# The Eclipsing ULX in NGC 3379

G. Fabbiano[1], D.-W. Kim[1], T. Fragos[2], V. Kalogera[2], A. R. King[3], L. Angelini[4], R. L. Davies[5], J. S. Gallagher[6], S. Pellegrini[7], G. Trinchieri[8], S. E. Zepf[9], A. Zezas[1]

[1] Harvard-Smithsonian Center for Astrophysics, 60 Garden St., Cambridge MA 02138; gfabbiano@cfa.harvard.edu, kim@cfa.harvard.edu, azezas@cfa.harvard.edu
[2] Northwestern University, Department of Physics and Astronomy, 2145 Sheridan Road, Evanston, IL 60208; tassosfragos@northwestern.edu, vicky@northwestern.edu
[3] University of Leicester, Leicester, LE1 7RH, UK; ark@star.le.ac.uk
[4] Laboratory for High Energy Astrophysics, NASA Goddard Space Flight Center, Code 660, Greenbelt, MD 20771; angelini@davide.gsfc.nasa.gov
[5] Denys Wilkinson Building, University of Oxford, Keble Road, Oxford; rld@astro.ox.ac.uk
[6] Astronomy Department, University of Wisconsin, 475 North Charter Street, Madison, WI 53706; jsg@astro.wisc.edu
[7] Dipartimento di Astronomia, Universita' di Bologna, Via Ranzani 1, 40127, Bologna, Italy; silvia.pellegrini@unibo.it
[8] NAF-Osservatorio Astronomico di Brera, via Brera 28, 20121 Milano, Italy; ginevra@brera.mi.astro.it
[9] Department of Physics and Astronomy, Michigan State University, East Lansing, MI 48824-2320; zepf@pa.msu.edu

ABSTRACT

We report recent *Chandra* observations of the ULX in the elliptical galaxy NGC3379 that clearly detect two flux variability cycles. Comparing these data with the *Chandra* observation of ~5yr ago, we measure a flux modulation with a period of ~12.6hr. Moreover, we find that the emission undergoes a correlated spectral modulation, becoming softer at low flux. We argue that our results establish this source as a ULX binary in NGC3379. Given the old stellar population of this galaxy, the ULX is likely to be a soft transient, however historical X-ray sampling suggests that the current on phase has lasted ~10yr. We discuss our results in terms of ADC and wind-feedback models. We constrain the donor mass and orbital period at the onset of mass transfer within 1.15-1.4$M_\odot$ and 12.5-16hr, respectively. The duration of the mass-transfer phase so far is probably ~1Gyr and the binary has been a soft X-ray transient throughout this time. These constraints are insensitive to the mass of the accretor.

1. INTRODUCTION

Ultra-luminous X-ray sources (ULXs) are non-nuclear X-ray sources observed in galaxies with X-ray luminosity in excess of $10^{39}$ erg s$^{-1}$. This observational definition translates in bolometric luminosities that exceed the Eddington luminosity of typical stellar black hole binaries (BH mass ~10M$_\odot$), making these sources possible candidates for intermediate mass black holes (IMBHs, with masses ~20M$_\odot$ and larger; see Long & Van Speybroeck 1983; Fabbiano 1989; Colbert & Ptak 2002). Alternatively, ULXs may represent a particular high-accretion state of normal X-ray binary evolution, with possibly anisotropic (King et al 2001) or super-Eddington emission (Begelman 2002), although other options have also been suggested [relativistic beaming (Koerding, Falke & Markoff 2002), young SN (Fabian & Terlevich 1996), pulsars (Perna & Stella 2004); see review in Fabbiano 2006].

ULXs tend to be found in galaxies with a high star-formation rate (e.g., the Antennae, Fabbiano et al 2001; the Cartwheel, King 2004), and are virtually missing in ellipticals, although in principle soft X-ray transients could appear as ULXs in these old populations (Piro & Bildsten 2002; King 2002). Irwin et al (2004) concludes that 'statistically' *Chandra* X-ray sources in E and S0 galaxies with $L_X > 2\times10^{39}$ erg s$^{-1}$ (~0.3-10 keV range) are likely to be background AGNs. However, a rare (although relatively faint) ULX was discovered in the inner regions of the elliptical galaxy NGC 3379 (Swartz et al 2004; David et al 2005). This source reached a peak luminosity of $3.5\times10^{39}$ erg s$^{-1}$ (corresponding to the Eddington luminosity of a 25 M$_\odot$ BH) and was found to be variable by a factor of ~50%, with a clear minimum observed during the 30ks *Chandra* exposure. Based on this time variability, David et al (2005) suggested that the ULX may be a binary system with orbital period of 8-10 hr and, based on this period and assuming that the secondary is filling its Roche lobe, estimated a mass of the companion star of ~1M$_\odot$.

As part of an ongoing *Chandra* legacy program, we are performing a series of deep monitoring observations of NGC 3379. The first of these new observations detected the ULX and showed that it varied both in flux and hardness ratio. In this letter we report these results that, in comparison with the archival *Chandra* data and a previous archival *ROSAT* observation, provide firm constraints on the nature of the ULX. We adopt a distance to NGC 3379 D = 10.57 Mpc throughout this paper, based on the surface brightness fluctuation analysis by Tonry et al. (2001). At the adopted distance, 1′ corresponds to ~3 kpc.

2. X-RAY OBSERVATIONS AND DATA ANALYSIS

NGC 3379 was observed for 85 ks on Jan. 23, 2006 with the *Chandra* Advanced CCD Imaging Spectrometer (ACIS) (obsid = 7073). The ACIS data were reduced as described in Kim & Fabbiano (2003) with a custom-made pipeline (XPIPE), specifically developed for the *Chandra* Multi-wavelength Project (ChaMP; Kim et al. 2004), using the most up-to-date calibrations (CALDB 3.2.1). Removal of background flares reduced the effective exposure time of CCD S3 to 80.2 ks.

NGC 3379 was previously observed with *Chandra* ACIS (obsid = 1587) for 30 ks on Feb. 13, 2001 (David et al. 2005). We have retrieved these data from the *Chandra* archive and we have reprocessed them to be consistent with the up-to-date calibration data (CALDB 3.2.1) used for the processing of the recent observation. These corrections take into account time-dependent variations of the ACIS response (QE, CTI, gain), and make the reprocessed data directly comparable with 7073. These data were then reduced with the same procedures used for the observation 7073.

Prior to *Chandra*, NGC 3379 had been observed with the *ROSAT* HRI for 24 ks on Feb. 8, 1996. David et al (2005) used this data to conclude that the ULX was at a comparable luminosity during that observation, although the source is confused with nearby X-ray emission. We retrieved the HRI data from the HEASARC *ROSAT* archive to independently estimate the X-ray flux of the ULX and we confirm the result of David et al. These three data sets were obtained at 5 years intervals, spanning 10 years, and provide a good baseline for a variability study (see §2.2).

Detailed data analysis was performed using the tools in CIAO v3.3.

2.1. Average fluxes and position of the ULX

The central part of the *Chandra* ACIS-S observation 7073, including the ULX, is shown in Figure 1. We measure an average flux of 2-3 x $10^{-13}$ erg s$^{-1}$ cm$^{-2}$ (see Table 1), and a position in the new *Chandra* observation of (RA, Dec) = (10 47 50.0, 12 34 56.8) in JD2000. Based on simulations developed for the ChaMP project (M. Kim et al 2006, in preparation), we estimated a centroid statistical error of 0.1". Typical *Chandra* absolute astrometry has 0.3'' uncertainty. This position is the same as that published by David et al (2005). The position of the ULX is 6.5" NE from the 2MASS position of the nucleus of NGC 3379 given by NED.

Its proximity to the center of NGC 3379 suggests that the ULX is indeed associated with NGC 3379. Based on the logN-logS relation determined by ChaMP (Kim et al. 2004), we calculate the chance for such a bright X-ray source ($f_X$ = 2 x $10^{-13}$ erg s$^{-1}$ cm$^{-2}$) to be a background source within the central region of a galaxy to be very small (only ~$10^{-5}$). We also compared the position of the ULX with a list of background galaxy positions, identified by Kundu, A. (2006, private communication) in the *Hubble* WFPC2 image which covers the center of NGC 3379 and includes the ULX. The nearest background galaxy is 4" away from the ULX (these comparison was done by referring the *HST* and *Chandra* positions to the same astrometric frame by matching globular cluster and X-ray source positions, see below). Moreover, the pattern of variability of the ULX (§2.2) suggests an X-ray binary.

We next discuss the possibility that this source may be a Galactic interloper. Given the flux, the X-ray luminosity would be too low for a Galactic X-ray binary with either a neutron star or a BH accretor, unless this system is in quiescence. Although the constant

X-ray flux over the 10 yr observing period would be consistent with this possibility, the high galactic latitude of the object (233.5, 57.6 deg.) places it far away from the Galactic plane, arguing against this hypothesis (White & van Paradjis 1996).

We therefore investigate the possibility of a binary with a white dwarf accretor, i.e. a foreground AM Her object. Roche geometry, together with the observed orbital period, implies that the companion has a mean density close to solar (see also David et al 2005). For stable mass transfer the companion must have a mass no greater than the accreting white dwarf, i.e. no more than about 1 $M_\odot$, requiring its radius to be about 1 $R_\odot$. Normal stars cannot have effective temperatures less than the Hayashi line, i.e. about 3000K, implying a companion luminosity > 0.06 $L_\odot$ and thus a minimum optical brightness for the system. We can compare this with the observed optical flux limit. This gives a lower limit on the distance, which we can compare with the distance to the 'edge' of the Milky Way along that line of sight. From the *Hubble* WFPC2 data we estimate a limit of V=25 mag for the optical counterpart to the ULX. This gives us a distance limit of 30 kpc. Given the Galactic coordinates of NGC 3379 this lower limit on the distance puts the source beyond the outer boundaries of the Milky Way. Moreover, the corresponding minimum X-ray luminosity, $4\times10^{34}$ erg s$^{-1}$, makes it extremely unlikely that this is an AM Her system. This luminosity is far higher than any known object of this type (e.g. Ramsay et al. 1994). The result is even stronger in that, in order for the white dwarf to be phase-locked to the orbital motion despite the wide separation implied by the detected period (see below), we require both a strong magnetic field and a low accretion rate, whereas the luminosity would imply an unprecedented high accretion rate.

The above considerations make a very strong case for the source to be a ULX in NGC 3379. We have further examined the possibility of an association with a globular cluster (GC) in NGC 3379. Two similarly luminous ULXs have been reported associated with GCs in the elliptical galaxy NGC 1399 (Angelini et al 2001). To this end, we have cross-correlated the positions of X-ray point sources in NGC 3379 with optical GCs discovered in the *Hubble* WFPT2 images by Kundu & Whitmore (2001). Using 6 LMXBs, which are clearly matched with GCs (within 1") and located near on-axis (within ~1'), but not very close the galaxy center (galactocentric distance = 10-60"), we correct for a systematic offset of 0.82'' (mostly in the RA direction). The remaining random offset is less than 0.3''. Independently, Kundu, Maccarone & Zepf (2006) determine GC-X-ray source matches in a number of galaxies including NGC 3379 and derive a relative astrometry with an uncertainty <0.4''. The nearest GC is found at ~ 1.35" SSE from the ULX. Given the 0.3-0.4'' accuracy of the relative astrometry and the centroid statistical error of 0.1", the discrepancy between the ULX and nearest GC position is ~3σ; we conclude that the ULX resides in the stellar field of NGC 3379.

2.2 Flux and Spectral Variability

During our new observations (7073), the ULX is variable both in flux and spectral hardness (fig 2). The flux light curve strongly suggests a period variability, covering two cycles. The peak X-ray luminosity (assuming the best fit power-law spectrum) is $L_X$ = 4-7 x $10^{39}$ erg s$^{-1}$ and the minimum luminosity is $L_X$ = 2-3 x $10^{39}$ erg s$^{-1}$. While the minimum is consistent with the measurement of David et al (2005), the maximum luminosity is slightly higher than that previously measured. Although this may be in part due to the limited phase coverage of the previous short observation, comparing the two observations, we find that the difference could be due to long-term variability (see fig. 3).

Combining the two *Chandra* observations spanning 5 years, we estimated the period of the variability. We used the CIAO application *sherpa* to fit simultaneously the two light curves with a sinusoidal curve plus a constant. We link the period and phase=0 epoch of two models to vary together, but allow the amplitude to vary independently since the mean flux may vary, as suggested by the comparison of our data with the previous *Chandra* observation (see above, fig. 3). To perform this fit, we binned the data so to have a minimum of 33 counts per bin, and we used $\chi^2$ statistics. We obtain best-fit reduced $\chi^2$ of 3.3 for 14 d.o.f. (7073), 0.6 for 12 d.o.f. (1587), and 2.0 for 28 d.o.f. for the joint fit. We notice that most of the residuals are due to two low bins and the last point in the 7073 light-curve. Excluding them, the reduced $\chi^2$ are 1.4 for 11 d.o.f. (7073) and 0.97 for 25 d.o.f. (joint). These discrepancies could well be due to our choice of model and to the presence of higher frequency components in the light curve, which are often observed in low-mass binaries (e.g. White et al 1995). We find that the best fit period for the observation 7073 (taken in 2006) is 12.3 ± 0.5 (1σ) hours or $12.7_{-1.1}^{+0.8}$ excluding the high residual points; as noticed by David et al (2005) the results for the observation 1587 (taken in 2001) are much less constrained, with a possible period of $9.2_{-2}^{+5}$ hours. The joint fit results in a period of 12.6 ± 0.3 hours, or 12.7±0.1 excluding the high residual points. The epoch corresponding to phase=0 is either (in TT) 98446798 sec in 2001 or 254390240 sec in 2006.

Although David et al. (2005) could not detect any variability of the spectral properties, fig. 2 shows that the spectral hardness also varies with a similar overall period, in the sense that the X-ray emission becomes harder at peak intensity, although there is substantial `flickering'. These hardness ratios are defined as H-S/H+S, where H and S are the counts in the 0.5-2.0 keV and the 2.0-8.0 keV band, respectively. The error bars are 1σ, in Gaussian approximation; using a Bayesian approach (Park et al 2006) that takes into account the asymmetric Poisson errors, the results are essentially consistent. Looking at the long-term behavior by comparing the average hardness ratio we measure in 7073 with that in 1587, in the same phase span, we find -0.48 ± 0.03 versus -0.64 ± 0.03 respectively. The average luminosities also differ by 23%. The lower luminosity emission is associated with a softer spectrum (see fig. 3).

We fitted the ACIS spectra of observation 7073 to models, for the entire observation and for two subsets obtained by extracting the data in two phase bins [observation time in TT sec = (254398400-254407400, 254438000-254454200) for the high flux and (254418200-254432600) for the low flux]. In both cases, the spectra were extracted from a circular region of 3" radius after excluding two nearby sources. The background was extracted locally from an annulus (10"-20"), but the background is negligible. The results are summarized in Table 1 for two choices of spectral model: a multi-color disk model, used to fit spectra of black hole binaries and ULXs (e.g. Makishima et al 2000), and a power-law. Using more complex composite methods does not improve the fit statistics. However, in cases where high signal to noise data are available, it is clear that complex spectral models are needed to fit ULX spectra (e.g., Miller et al 2003; Goad et al 2006). These results should only be considered indicative of the ULX spectral parameters, and a full fledged spectral analysis requiring significant higher signal to noise will be postponed to a future time, after the completion of our monitoring program. In both cases, we performed fits freezing the absorption column ($N_H$) to the line of sight value of 2.78 x $10^{20}$cm$^{-2}$ (from COLDEN), and also letting this parameter free to vary. In all cases, we obtain acceptable values of $\chi^2$, but the range of $N_H$ is largely undetermined; we only list the results for line of sight $N_H$ in Table 1.

Table 1.

| Data Set | Net Counts ± Error (1σ) | Model | $T_{in}$ (keV) or Γ | $\chi^2$/d.o.f | $f_{X,\ 0.3-10keV}$ ($10^{-13}$ erg s$^{-1}$ cm$^{-2}$) (unabsorbed) | $L_{X,\ 0.3-10keV}$ ($10^{39}$ erg s$^{-1}$) |
|---|---|---|---|---|---|---|
| 7073 all | 2462.4±49.7 | MCD | $1.41_{-0.08}^{+0.09}$ | 61.3/89 | 2.16 | 3.0 |
| | | Power-law | $1.26_{-0.04}^{+0.04}$ | 68.4/89 | 3.35 | 4.7 |
| 7073 hi | 888.8±29.9 | MCD | $1.88_{-0.22}^{+0.29}$ | 12.7/35 | 3.16 | 4.4 |
| | | Power-law | $1.09_{-0.-07}^{+0.07}$ | 14.6/35 | 4.63 | 6.5 |
| 7073 lo | 310.1±17.7 | MCD | $1.07_{-0.24}^{+0.12}$ | 3.9/11 | 1.38 | 1.9 |
| | | Power-law | $1.62_{-0.15}^{+0.15}$ | 2.6/11 | 2.21 | 3.1 |
| 01587 all | 734.6±27.2 | MCD | $0.98_{-0.47}^{+0.014}$ | 25.1/29 | 1.17 | 1.6 |
| | | Power-law | $1.79_{-0.07}^{+0.08}$ | 25.1/29 | 1.65 | 2.3 |
| 7073 (1587)* | 758.9±27.6 | MCD | $1.13_{-0.19}^{+0.06}$ | 21.6/30 | 1.60 | 2.2 |
| | | Power-law | $1.50_{-0.08}^{+0.08}$ | 20.6/30 | 2.49 | 3.5 |

* This spectrum was extracted from the 7073 observation at the same phase with the 1587 observation, i.e., during phase = 0.36 – 1.02. The count rate for 1587 was corrected for the ACIS QE variation so that it could be compared with that of 7073.

In general, the results of table 1 follow the hardness ratio results: the MCD inner temperature of the accretion disk $T_{in}$ appears larger (harder spectrum) in the high flux. Similarly, the power-law tends to be flatter in the high flux. The overall spectrum is softer in the observation 1587, where the overall source flux was also lower, when compared the 7073 observation extracted at the same phase with the 1587 observation (phase = 0.36 – 1.02). In all cases, uncertainties are at a 1σ confidence level for 1 interesting parameter.

3. DISCUSSION

We have observed a full cycle of variability in the ULX in NGC 3379 and we conclude that this source is a periodic variable, with a period of ~12.5 hr, by fitting a sinusoidal model to the entire set of *Chandra* observations. Although the shape of the light curve is somewhat uncertain, given the error bars, and possibly not sinusoidal, the minima are well defined. Given the 5 yr baseline, we believe that our period determination is robust. This variability strongly suggests that the ULX is an X-ray binary, and that the variability is orbital. We also discovered that the spectral properties of the emission seem to undergo a correlated (although noisier) variability cycle, with softer emission observed at the minima of the light curve. Comparing our recent observation with the previous *Chandra* observation discussed in David et al (2005) we have also found evidence of minor (~23%) long-term flux variability over a 5 yr span, in the sense that the ULX was dimmer in the year 2001 than in 2006, considering a comparable phase of the light curve. We also measure a softer X-ray hardness ratio (and spectrum) in the data of 5 yrs ago. Since the ULX is also visible in the archival *ROSAT* data (see also David et al 2005), we conclude that the ULX may have been steadily emitting for ~10 yr. If we assume an isotropic emission at the Eddington limit, the peak $L_X$ corresponds to $M_{BH} = 32$ M$_\odot$.

This ULX appears not to be associated with a globular cluster. Although one could argue that the ULX may have been ejected from the neighboring cluster, or that the parent GC may have evaporated or been tidally disrupted (the ULX is only ~340pc from the nucleus of NGC 3379), its large luminosity (a factor of ~10 higher than the Eddington limit for a neutron star accreting He-rich material, see Shakura & Sunyaev 1973) would be hard to explain in the GC formation scenario proposed by Bildsten & Deloye (2004) that LMXB in ellipticals may be ultra-compacts with NS accretors formed in globular clusters. BH binaries in the field (see Ivanova & Kalogera 2006) may instead be very luminous sources, because of a higher Eddington limit during transient outbursts. Transient behavior here is very likely (see below) because of the relatively long binary period and the small donor mass, given the old age of the stellar population (see King, Kolb & Burderi, 1996).

Since the stellar population of NGC 3379 is old with an age of 9-10 Gyr (see the recent *Sauron* study of line strength maps, Kuntschner et al 2006; we confirm these result for the position of the ULX with a new look at the *Sauron* data), the donor star feeding the compact object must be of low mass ~1M$_\odot$. Thus the most likely interpretation is that the ULX is a soft X-ray transient in outburst (Piro & Bildsten 2002; King 2002). We note that the transient character of the source is expected for both stellar-mass and intermediate mass black holes. The metallicity for NGC3379 is estimated to be a factor of 1.5-2 higher than solar metallicity (Z~0.03-0.04; Terlevich & Forbes 2002; Trager et al 2000; Thomas et al 2005). Assuming that the binary orbital period is 12.5hr at Roche-lobe overflow, we run stellar evolution calculations using an up-to-date stellar evolution code described in detail in Podsiadlowski et al. (2002), Ivanova et al. (2003), and Kalogera et al. (2004), and Ivanova & Taam (2004). We find that the donor mass is very narrowly constrained in the range 1.1-1.15M$_\odot$, given a galaxy age in the range 8-10Gyr. This constraint is highly insensitive to the accretor's mass (values of 10,100, and 1000 M$_\odot$ were examined). We further examine mass-transfer simulations for such donor masses, and, as expected, the calculated mass-transfer rates are indeed lower than the critical values for transient behavior (critical rates derived by Dubus et al. 1999 were used).

However, most probably mass transfer did not start at the currently measured period of 12.5hr. It is possible that it started at either (i) a longer period for a donor star more massive than $1.15M_\odot$ and the orbit has been shrinking due to magnetic braking and gravitational radiation, or (ii) at a shorter period for a donor star less massive than $1.1M_\odot$ and the orbit has been expanding due to the donor's nuclear evolution. Our mass-transfer simulations indicate that the latter hypothesis can be excluded because, for magnetic braking strengths typically used in the literature (Rappaport et al. 1983; Ivanova & Taam 2003), radial expansion due to nuclear evolution of stars less massive than $1.1M_\odot$ cannot overtake magnetic braking and drive orbital expansion. Therefore mass transfer for the observed system must have started at an orbital period longer than 12.5hr; the initial donor mass must have been higher than $1.15M_\odot$, but low enough so that a significant convective envelope mass be present, and consequently magnetic braking be operational and strong enough to drive orbital contraction.

The requirement for the presence of a partially convective envelope constrains the initial donor mass to be lower than 1.3 and 1.4 $M_\odot$, for Z=0.03 and 0.04, respectively (this limit is $1.25M_\odot$ for solar metallicity). We have examined a set of mass-transfer calculations for initial donor masses up to 1.5 $M_\odot$ and initial orbital periods up to 18hr. For Z=0.03 and 0.04, we find that the properties of the ULX at the onset of mass transfer are further constrained to lie in the range 1.15- $1.4M_\odot$ for the donor mass and 12.5-16hr for the orbital period, for any accretor mass. In these ranges the maximum orbital period allowed decreases as the donor mass increases.

For properties outside these ranges the binary expands instead of contracting either because the donor mass no longer has an outer convective envelope or the donor's radial expansion due to nuclear evolution dominates over magnetic braking. With these tight constraints on the binary properties at the onset of mass transfer we are also able to constrain the duration of the mass transfer phase (i.e., the time from the mass-transfer onset until the orbital period reaches the current measurement of 12.5hr) to shorter than 1Gyr (down to just about 10Myr, if mass transfer starts very close to 12.5hr). We note that for the highest donor masses in the accepted range (1.3-$1.4M_\odot$) the total age of the system is 5-6Gyr, somewhat shorter than the galaxy age estimated from the *Sauron* study; such a difference is acceptable, given the uncertainties associated with the galaxy age estimate. If we do not allow for this difference, then the donor mass would be even more narrowly constrained (1.15-$1.25M_\odot$). Last we note that for the mass-transfer sequences that satisfy all these constraints, the mass transfer rate remains below the critical rate for transient behavior throughout, consistent with more general theoretical expectations that bright sources in ellipticals should be transient (Piro & Bildsten 2002; King 2002; Ivanova & Kalogera 2006).

The maximum length of the outburst is set by the geometric size of the accretion disc: if the black hole mass is ~30 $M_\odot$, a period of ~12.5 hr implies an orbital separation of ~ $6\times10^{11}$ $(M/30M_\odot)^{1/3}$cm, where M ~ 30 $M_\odot$ is the total binary mass. This suggests a total disc mass before the outburst of ~2 $\times10^{27}$ g, depending on the disc viscosity [King & Ritter 1998, eqn (8)]. If the disc mass is consumed at the rate indicated by the ULX luminosity, the outburst could last ~ 7 years. However the viscosity is highly uncertain, and in addition the source may well be super-Eddington, i.e. actually consuming the mass faster than this rate. The length of the outburst could thus vary significantly from this estimate. Empirically one might expect it to be longer than that of A0620-00 (few 100 days; see Tanaka & Lewin 1995) and shorter than that of GRS 1915+105 (10 years and still going; see Fender & Belloni 2004) as the orbital period is between the two (6.5 hours and 33 days respectively).

The source is never totally eclipsed, a behavior reminiscent of Galactic LMXBs (see White et al 1995), but narrow eclipses may be hidden by the relatively low signal to noise ratio of the data. The light curve expected for a point X-ray source eclipsed by a companion would be flat (plus aperiodic modulations) apart from a very narrow total eclipse. A much more usual pattern of orbital modulation for LMXBs is shown by the so-called dippers and ADC (Accretion Disk Corona) sources. In these binaries, the X-ray source is slightly extended, and has a structure fixed in the orbital frame, creating a more complex periodic light curve. These sources are believed to have matter close to the accretor, which probably scatters the X-rays and makes an extended X-ray source component. In ADC sources the light curve does not have a deep narrow minimum, while in dippers it does (see e.g. Frank, King & Raine 2002, pp 106, 107). The two classes differ only in inclination angle, the dippers having lower inclination than the ADCs.

The hardness ratio behavior seen for the NGC 3379 ULX seems to be opposite from what one usually sees in dippers. In ADCs we do not see the central point X-ray source, but only the scattered X-rays. Although these sources typically do not present an orbital modulation of the hardness ratio, the corona may have an uneven structure or there may be some intervening material responsible for partial absorption of the X-ray s from the corona and possible re-emission. Galactic ADC sources are intrinsically faint, because for low scattering optical depths $\tau$, the scattered X-rays have luminosity of order $\tau$ times the central point source luminosity. Thus one usually expects the unseen central point source to be much brighter than the detected luminosity. In known ADCs, comparison of optical/X-ray ratios with face-on sources suggests that the central source is 10 – 100 times brighter than the observed (scattered) luminosity (White & Holt 1982; see Mason 1986 for a comparison of different X-ray binaries). In our case this would mean a true luminosity significantly higher than the $\sim 4 \times 10^{39}$ erg s$^{-1}$ we infer from the observed flux. However for accretion rates near or above the Eddington rate, as is likely for our object, we must have $\tau > 1$ (King & Pounds 2003), so the extended source luminosity may be comparable to the (unseen) point source. Our object may thus give us insight into what happens when accretion is at or above the Eddington value, and indeed into the nature of the ULX phenomenon in general.

If the source is not an ADC, a different model that may apply is that of wind-feedback (e.g., Basko et al 1977, and references therein). In this model, illumination of the companion by X-rays from the compact object causes heating of the companion and promotes stronger stellar winds; at perigee the wind will have the effect to make the emission harder since it will act as an absorber. This will model would be consistent with the observed lack of total eclipses and the smooth modulation of the light curve. A better-defined X-ray light curve and spectral light-curve are needed to firmly establish the nature of the source.

## 4. CONCLUSIONS

Our recent *Chandra* observations of the elliptical galaxy NGC 3379, in conjunction with archival *Chandra* and *ROSAT* data (see David et al 2005) have led to the measure of a 12.6±0.3 hr period in the variability of the luminous ULX present in this galaxy (Swartz et al 2004). We also found correlated spectral variability, with the emission becoming softer in the minima of the light curve. Including our new data, this ULX has been detected with a similar average luminosity over a 10 yr time span. Given the metallicity and the old age of the stellar population in NGC 3379 and assuming that the ~12.5 hr period is that of the orbit, we are able to constrain the donor mass and orbital period at the onset of mass transfer within 1.15-1.4$M_\odot$ and 12.5-17hr, respectively. The duration of the mass-transfer phase so far is probably ~1Gyr (although we cannot exclude that it is much shorter about ~10Myr) and the binary has been a soft X-ray transient throughout this time. These constraints appear to be quite insensitive to the assumed accretor mass (10, 100, 1000 $M_\odot$). The light curve and spectral behavior may be consistent with an ADC binary, although in this case the intrinsic luminosity may be significantly higher than suggested by the detected flux. This source may thus give us some real insight into what happens in super-Eddington accretion and thus possibly into the ULX phenomenon in general. However, the spectral modulation of the light curve is not typical of ADC sources and may alternatively suggest a wind-feedback model (e.g. Basko et al 1977). To really constrain the nature of this ULX, significantly better light curve and spectra are needed, for a more accurate comparison with well studied Galactic binaries. If the ULX continues to shine in the upcoming year, our *Chandra* legacy program will provide the needed data.

## ACKNOWLEDGMENTS

This work was supported by the Chandra GO grant GO6-7079A (PI Fabbiano) and subcontract GO6-7079B (PI Kalogera). D.-W. Kim acknowledges support from NASA contract NAS8-39073 (CXC); A. Zezas acknowledges support from NASA LTSA grant NAG5-13056. The data analysis was supported by the *CXC* CIAO software and CALDB. We have used the NASA NED and ADS facilities, and have extracted archival data from the NASA HEASARC and *Chandra* archives. We thank Arunav Kundu for providing positions of globular clusters from his *Hubble* observations, Davor Krajnovic for a re-analysis of the *Sauron* data, and Natasha Ivanova for allowing us to use her stellar evolution and mass transfer code.

**Figure Captions**

Figure 1. *Chandra* X-ray image near the center of NGC 3379 (green square from 2MASS) and ULX (red arrow), nearest GC (blue cross) and nearest background galaxy (green cross) marked.

Figure 2: Light curve of the ULX. The count rate (top) and hardness ratio (bottom) are plotted again time (TT). In both plots, the black curve is the best-fit sinusoidal to the count rate light curve (see text). The red horizontal bars at the top of the figure indicate time intervals used for the high flux spectrum, the blue bar indicates the interval used for the low-state spectrum (see Table 1) and the cyan bar indicates the phase covered during the 1587 observation (see Figure 3).

Figure 3: same as Figure 2, but with the previous observation 1587 taken 5 years ago. The data have been corrected for the temporal variations of the ACIS quantum efficiency. The sinusoidal curve is the best-fit sinusoidal from the entire ACIS data set (see text).

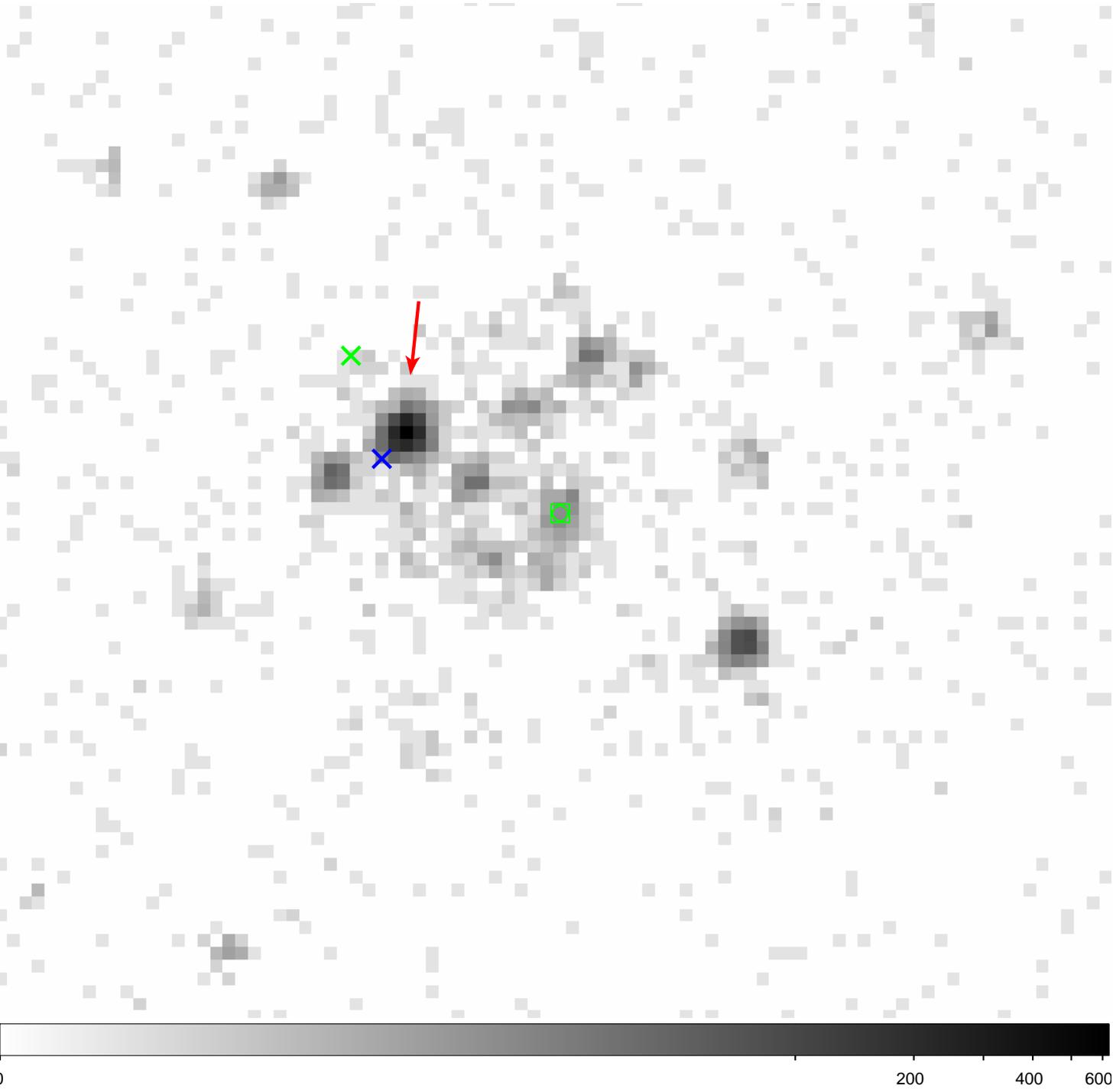

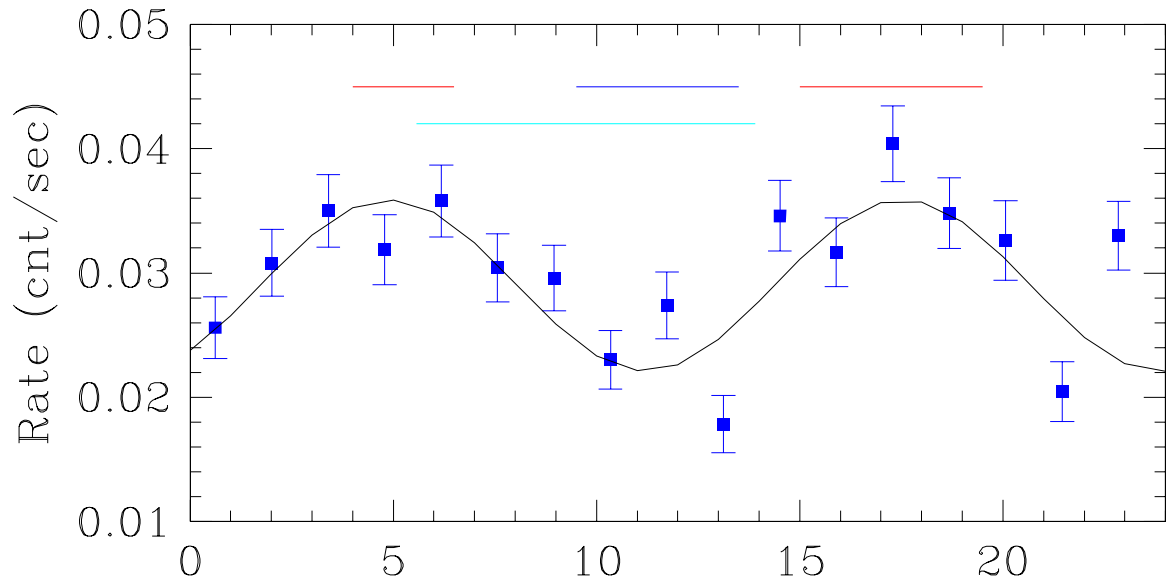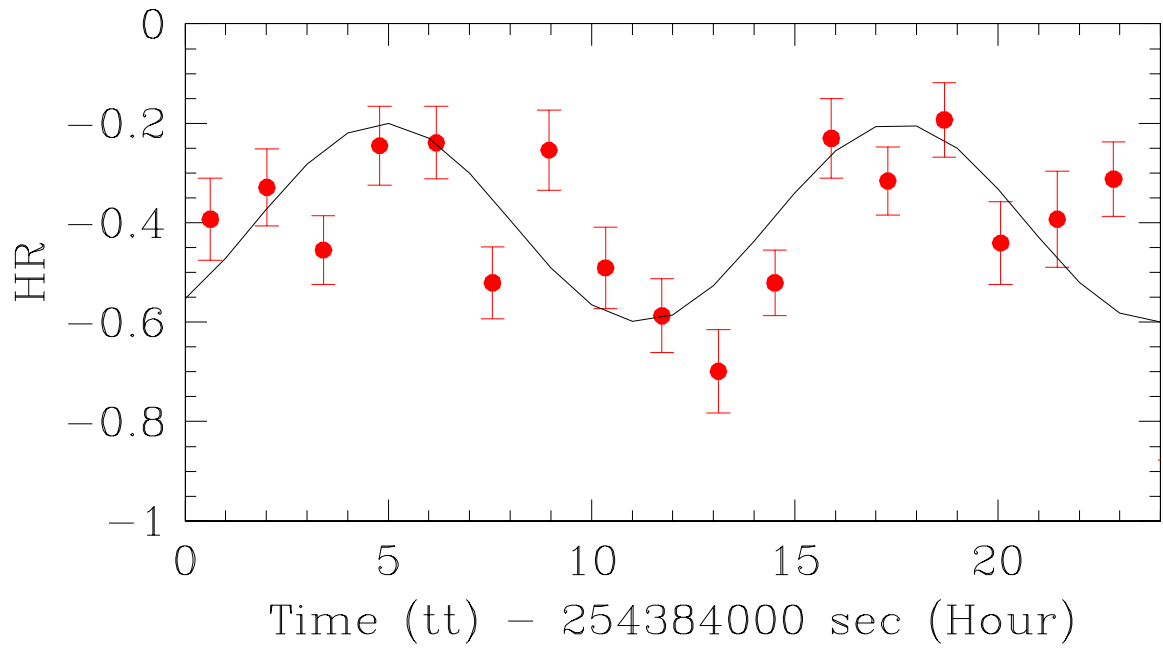

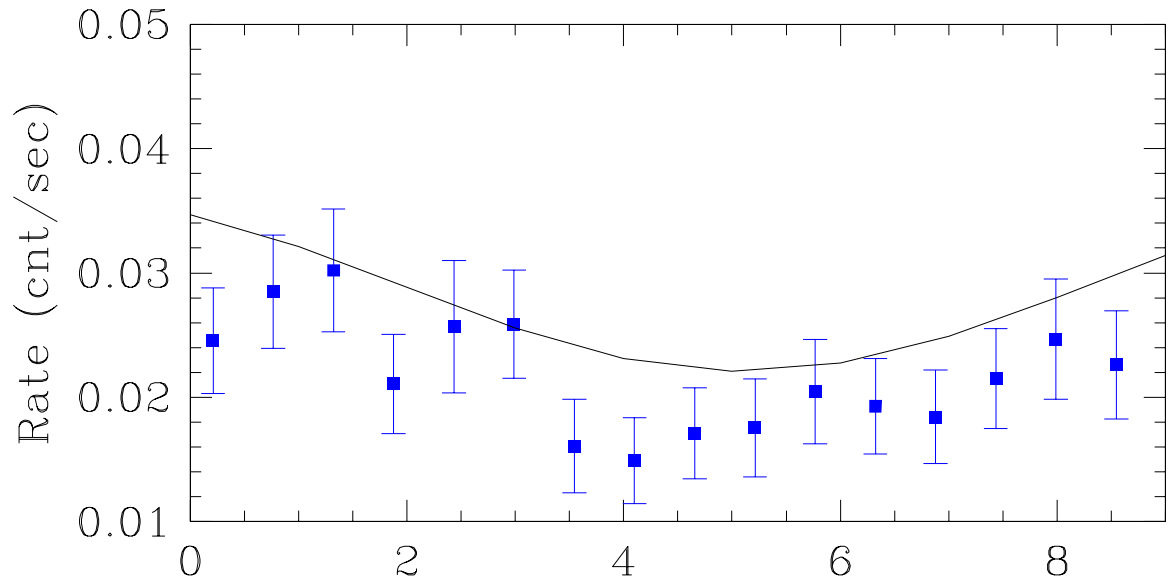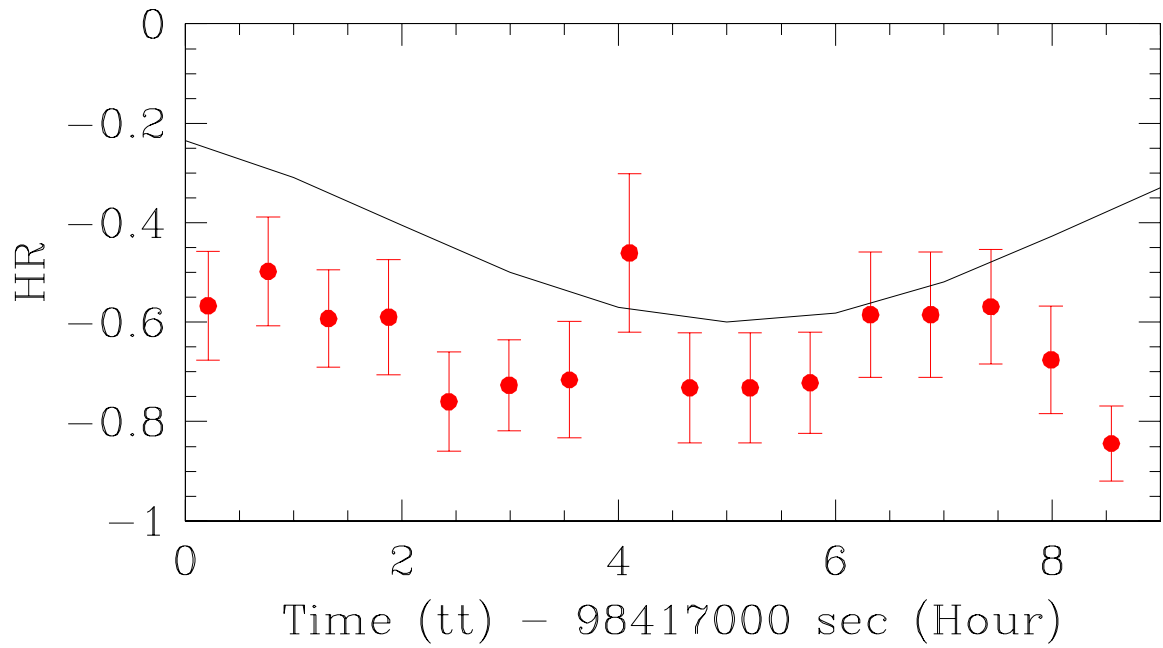